\documentclass[aps,letterpaper,twocolumn,pra,nofootinbib,floatfix]{revtex4}
\usepackage{graphicx}
\usepackage{amsmath}
\usepackage{amsfonts}
\usepackage{amssymb}
\usepackage[scanall]{psfrag}

\begin{document}
\newcommand{\of}[1]{\left( #1 \right)}
\newcommand{\sqof}[1]{\left[ #1 \right]}
\newcommand{\abs}[1]{\left| #1 \right|}
\newcommand{\avg}[1]{\left< #1 \right>}
\newcommand{\cuof}[1]{\left \{ #1 \right \} }
\newcommand{\bx}{\mathbf{x}}
\newcommand{\by}{\mathbf{y}}
\newcommand{\bk}{\mathbf{k}}
\newcommand{\bp}{\mathbf{p}}
\newcommand{\bl}{\mathbf{l}}
\newcommand{\bq}{\mathbf{q}}
\newcommand{\br}{\mathbf{r}}
\newcommand{\Uhat}{\widehat{U}}
\newcommand{\up}{\uparrow}
\newcommand{\down}{\downarrow}
\title{Even-Odd Correlation Functions on an Optical Lattice}
\author{Eliot Kapit} 
\email{ek359@cornell.edu}
\affiliation{Laboratory of Atomic and Solid State Physics, Cornell University, Ithaca, NY}
\author{Erich Mueller}
\affiliation{Laboratory of Atomic and Solid State Physics, Cornell University, Ithaca, NY}

\begin{abstract}
We study how different many body states appear in a quantum gas microscope, such as the one developed at Harvard [Bakr et al. Nature 462, 74 (2009)], where the site-resolved parity of the atom number is imaged. We calculate the spatial correlations of the microscope images, corresponding to the correlation function of the parity of the number of atoms at each site. We produce analytic results for a number of well-known models: noninteracting bosons, the large U Bose-Hubbard model, and noninteracting fermions.  We find that these parity correlations tend to be less strong than density-density correlations, but they carry similar information.
\end{abstract}
\maketitle
\section{Introduction}

Bakr \textit{et al.}\cite{greiner} recently reported site resolved images of atoms in optical lattices.
Their tool, the quantum gas microscope, works by rapidly tuning the lattice near resonance with an atomic transition: the resulting increase of the atomic confinement freezes the atoms in place.  Simultaneously, the sample is bathed with a separate near-resonant beam of light, causing the atoms to fluoresce, and allowing individual atoms  to be observed. While this technique enables single-atom imaging \cite{sherson}, it has a side effect that sites with an even number of atoms are quickly emptied by light-assisted collisions, and appear dark.  The same effect causes all but one atom to be lost at sites with an odd number of atoms.  Hence, the quantum gas microscope images map out the atomic number parity $P_{\bx} \equiv ({1 - \of{-1}^{n_{\bx}} })/{2}$, where $n_{\bx}$ is the density at site $\bx$.  Here we investigate correlations 
\begin{equation}\label{defd}
D_{\bx \by} \equiv \avg{ P_{\bx} P_{\by} } - \avg{P_{\bx}} \avg{P_{\by} }, 
\end{equation}
which can be measured by taking autocorrelations of the quantum gas microscope images.  We compare this correlation function to the more typical density-density correlations, $\avg{n_{\bx} n_{\by}}-\avg{n_{\bx}}\avg{n_{\by}}$.  Measurements of density correlations
have been quite valuable for learning about cold gas systems \cite{greinerjin,bloch2, hofferberth, spielman, inguscio, hadzibabic}, and have future applications \cite{theorycor}.  Of particular interest is a protocol introduced by Zhou and Ho\cite{ho} which allows one to accurately determine the temperature of a cold atomic system by averaging the density and density fluctuations of the system across the entire harmonic trapping region.
%

We consider several different cases:
(1) free or weakly interacting bosons, (2)  the large $U$ Bose-Hubbard model, and (3) free fermions. While $D_{\bx \by}$ is much harder to interpret than the density-density correlator  $\avg{n_{\bx} n_{\by}}-\avg{n_{\bx}}\avg{n_{\by}}$, we find that it is a very useful probe.  In particular, 
there is much information contained in these correlation functions which is not simply found in the mean $\avg{P_{\bx}}$.  Although these are all simple models, our results are nontrivial.  Calculating these correlations can be difficult, even for noninteracting bosons.  Furthermore, as all of these systems have been realized, our results are relevant to experiments.

There are many interesting systems (such as the t-J model), where the occupation on a given site is restricted to be 1 or 0.  For those systems there is a one-to-one correspondence between the parity correlations and the density-density correlations.  This makes these models extremely well suited to experiments of this type, but we will not treat them here since their correlation functions can be calculated through standard means.

Additionally, one may be able to use knowledge of the underlying physics to directly extract $n_i$ from the measurement $P_i$. For example, in the large $U$ Hubbard model number fluctuations are sufficiently attenuated that at any point in space the density is likely to take on only one of two values. Depending on what quantities one is interested in, studying these $n_i$ images may be more fruitful than the analysis presented here.

Throughout we will consider a uniform system.   We envision using a local density approximation (LDA) to apply our results to a trapped gas.  For the LDA to be valid we need two conditions.  First the confinement of the cloud must be sufficiently weak that the system is locally homogeneous.  Typically this condition is satisfied if the harmonic trapping frequency is small compared to the tunneling rate, $\omega_{c} \ll t$.   Second, we require that the two points in our correlation function,  $\bx$ and $\by$, are separated by a distance much smaller than the size of the cloud.  Thus there is an upper limit on the length scale over which we can measure $D_{\bx \by}$ \cite{hooley}.  If $D_{\bx \by}$ decays sufficiently rapidly, then we can completely characterize this function by experiments on a finite cloud.

The remainder of this paper is organized as follows.  In section~\ref{bose} we give results for weakly interacting bosons.  In section~\ref{largeU} we consider a bosonic Mott insulator.  In section~\ref{ferm} we discuss the two-component Fermi gas.

\section{Bosons: ideal and weakly interacting}\label{bose}

The most technically challenging of our calculations will be for the case of finite temperature noninteracting Bosons.  We will also explain how weak interactions can be included, but will not give detailed expressions.  In these systems the correlator $D_{\bx \by}$ encodes information about the density-density correlation length and number squeezing.  Our central result will be analytic expressions for  $D_{\bx \by}$ which are exact in the thermodynamic limit.

\subsection{Bose-Einstein Condensate in the ideal gas}\label{ideal}
\begin{figure}

\includegraphics[width=1.5in]{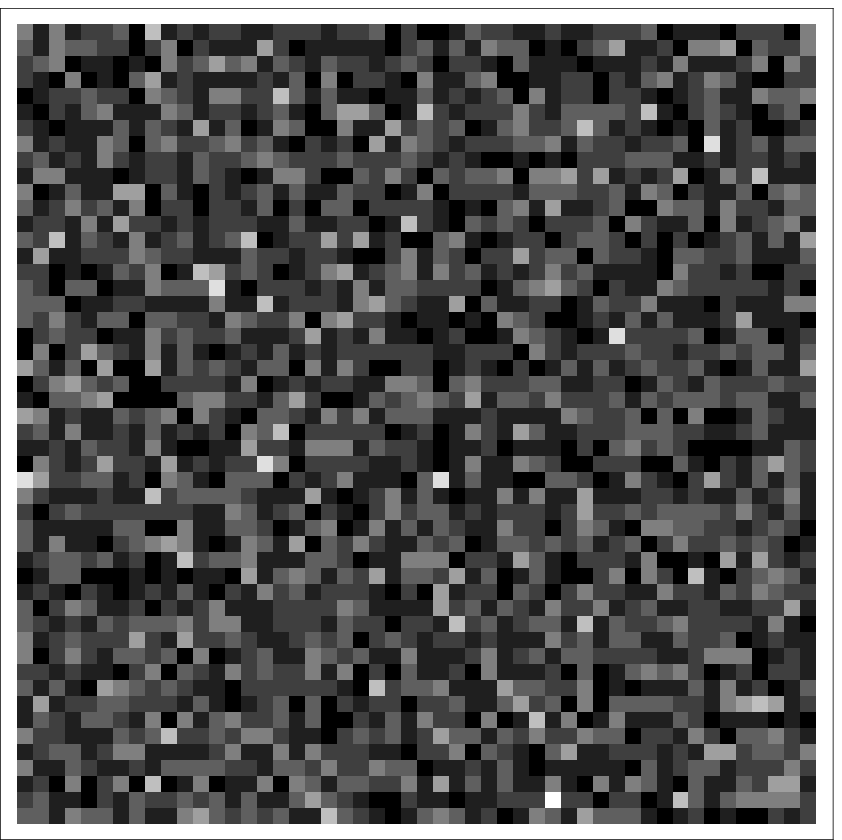}
\includegraphics[width=1.5in]{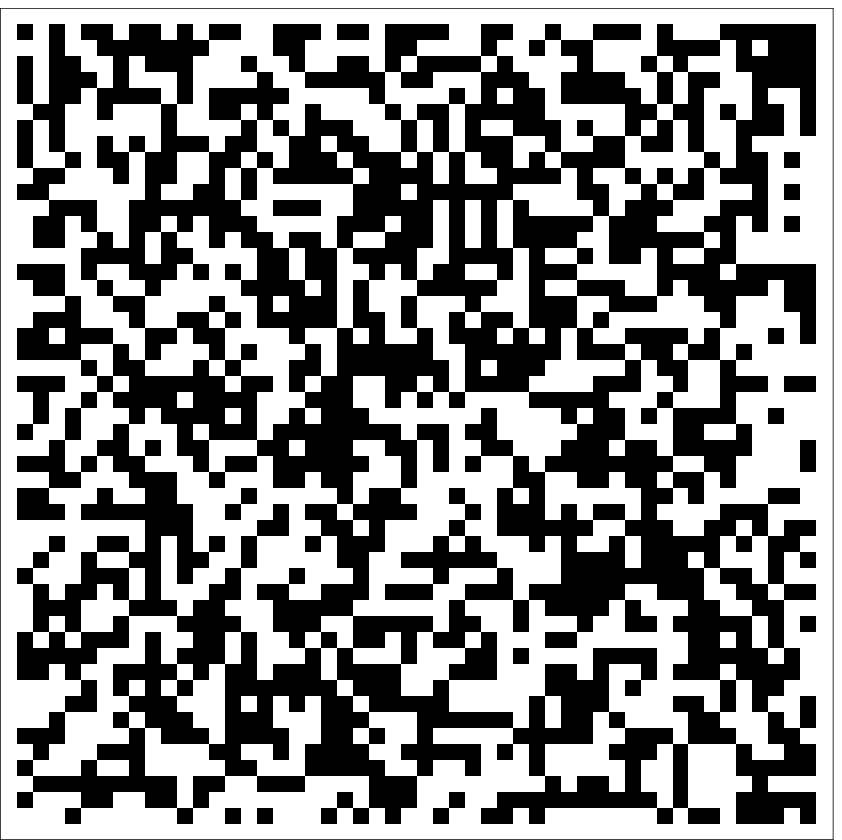}

\caption{Characteristic images of the density (left) and parity (right) of
a homogeneous noninteracting BEC in an optical lattice
at zero temperature with average density $\avg{n}=2$. 
Left: brighter colors correspond to higher density.  Right: white sites correspond to odd numbers of particles, and black corresponds to even.}\label{idealfig}

\end{figure}

The simplest limit to consider is a zero-temperature ideal (noninteracting) Bose gas.  As is well described in the literature \cite{bloch}, the number occupation on different sites will then be Poissonian, and uncorrelated, yielding $D_{\bx \by}=0$. The expectation value of the single site atom number parity operator is
\begin{eqnarray}
\avg{ P_{\bx}}&=&(1-e^{-2\avg{n_{\bx}}})/2.
\end{eqnarray}
Figure~\ref{idealfig} shows the expected quantum gas microscope images.

\subsection{Finite temperature ideal Bose gas}\label{finite}
Scaling energies by the temperature, a gas of noninteracting bosons on an optical lattice is described by the Hamiltonian
\begin{eqnarray}
H = \sum_{\bk} \of{ \epsilon_{\bk} - \mu } c_{\bk}^{\dagger} c_{\bk},
\end{eqnarray}
where $c_{\bk}$ annihilates a boson with quasimomentum $\bk$.
For calculating the properties of this system we
will work on a finite lattice of $N$ sites, and it will be understood that all calculations  are to be evaluated in the limit $N \rightarrow \infty$. 
Our results will apply to a wide varieties of lattices with a range of different parameters.
We shall require that the dispersion $\epsilon_{\bk}$ is even, which is generically true unless time reversal symmetry is broken.  In this subsection we will further require that $\of{\epsilon_{\bk} - \mu}$ is positive definite, implying that the gas is noncondensed.  Under these conditions we will be able to calculate $D_{\bx \by}$ as a function of $\mu$ and $T$.  In section~\ref{sf} we will analyze the superfluid state.

We perform our calculation in a momentum occupation basis:
\begin{equation}
\left| \{n_k\} \right> = \sqrt{ \prod_{i = 1}^{M} \frac{1}{n_{i} !} } \of{c_{\bk_{1}}^{\dagger}}^{n_{1}} \of{c_{\bk_{1}}^{\dagger}}^{n_{2}} ... \of{c_{\bk_{1}}^{\dagger}}^{n_{M}} \left| 0 \right>,
\end{equation}
where $M$ is the number of distinct occupied momentum states.
The partition function can be written as
\begin{equation}
Z = \sum_{\{n_k\} } \left< \{n_k\}  \right| e^{-H} \left| \{n_k\}  \right>,
\end{equation}
were
the sum is over all possible occupations of the momentum eigenstates. 

\subsubsection{Single Site Parity Expectation Value}
We begin by calculating the expectation value
\begin{widetext}
\begin{eqnarray}
\avg{ \of{-1}^{n_{\bx}} } 
= Z^{-1} \sum_{\{n_k\}} \prod_{i = 1}^{M} \frac{1}{n_{i} !} \left< 0 \right| \of{c_{\bk_{1}}}^{n_{1}} 
\cdots \of{c_{\bk_{M}}}^{n_{M}} 
 \of{-1}^{n_{\bx}} \of{c_{\bk_{1}}^{\dagger}}^{n_{1}}  
\cdots \of{c_{\bk_{M}}^{\dagger}}^{n_{M}} \left| 0 \right> e^{-\beta \sum_k E_k n_k},
\end{eqnarray}
where we define $E_k=(\epsilon_k-\mu)$ and $\beta = 1/T$.
We define our fourier transform conventions as:
\begin{eqnarray}
c_{\bk} = \frac{1}{\sqrt{N}} \sum_{\bx} c_{\bx} e^{i \bk \cdot \bx}, \quad c_{\bk}^{\dagger} = \frac{1}{\sqrt{N}} \sum_{\bk} c_{\bx}^{\dagger} e^{-i \bk \cdot \bx}.
\end{eqnarray}
To evaluate the expectation value, we commute the operator $\of{-1}^{n_{\bx}}$ through all the $c^{\dagger}$ operators to the vacuum state, where it evaluates to 1. In the position basis, $ \of{-1}^{n_{\bx}} c_{\by}^{\dagger} = \of{ 1 - 2 \delta_{\bx \by} } c_{\by}^{\dagger} \of{-1}^{n_{\bx}}$. In the momentum basis, this yields 
\begin{eqnarray}\label{deflambda}
\of{-1}^{n_{\bx}} c_{\bk}^{\dagger} &=& \of{ c_{\bk}^{\dagger} - \frac{2\lambda}{\sqrt{N}} c_{\bx}^{\dagger} e^{-i \bk \cdot \bx} } \of{-1}^{n_{\bx}}
 \equiv d_{\bk}^{\dagger} \of{-1}^{n_{\bx}}\\\nonumber
\sqof{c_{\bp},d_{\bk}^{\dagger}} &=& \delta_{\bp \bk} - \frac{2 \lambda}{N} e^{i \of{\bk - \bp} \cdot \bx},
\end{eqnarray}
where $\lambda=1$ will be used as a formal expansion parameter.  Using these operators we write
\begin{eqnarray}\label{bigexp}
\avg{ \of{-1}^{n_{\bx}} } = Z^{-1} \sum_{\{n_k\}} \prod_{i = 1}^{M} \frac{1}{n_{i} !} \left< 0 \right| \of{c_{\bk_{1}}}^{n_{1}}
\cdots
\of{c_{\bk_{M}}}^{n_{M}} \of{d_{\bk_{1}}^{\dagger}}^{n_{1}}
\cdots
\of{d_{\bk_{M}}^{\dagger}}^{n_{M}} \left| 0 \right> e^{-\beta \sum_k E_k n_k}.
\end{eqnarray}
\end{widetext}
This expectation value can be calculated as a product of vacuum expectation values via Wick's theorem.  We construct a formal power series in $\lambda$ [introduced in Eq.~(\ref{deflambda})]: $\langle(-1)^{n}\rangle=\sum_m F\of{m} \lambda^m$.  We construct this expansion by introducing the contractions
\begin{eqnarray}
\overline{c_{\bk} d_{\bp}^{\dagger}} \equiv \sqof{ c_{\bk}, d_{\bp}^\dagger} -\delta_{\bk\bp}
=-\frac{2\lambda}{N}e^{i (\bk-\bp)\cdot\bx}.
\end{eqnarray}  
Mathematically, $F\of{m}$ counts the contribution to $\langle(-1)^{n}\rangle$ in which there are $m$ such contractions.
Each of these contractions caries a factor of $1/N$, where $N$ in the number of sites, but we will find an additional combinatorial factor of ${\cal O}(N^m)$ which offsets this.  Thus,  when $\lambda=1$ (as is physical), each term in this expansion will be of comparable size and we will need to sum the entire series to produce a useful result.



%


We will begin by explicitly constructing $F(m)$ for $m=0,1,2$, and then will provide an argument for the general term. The leading term is $F(0)=1$, as the sum in Eq.~(\ref{bigexp}) is then simply the partition function.    

Next,  $F(1)$ comes from a single contraction of equal momenta, $\overline{c_{\bp} d_{\bp}^{\dagger}}$. Contractions which involve two distinct momenta such as $\overline{c_{\bp} d_{\bk}^{\dagger}}$ will leave unequal numbers of $c_{\bp}$ and $d_{\bk}^{\dagger}$ operators in Eq.~(\ref{bigexp}). The expectation value of these will vanish unless an additional off-diagonal contraction is made, contributing another power of $\lambda$.  
   Let $\bp$ be the momentum involved in the contraction $\overline{c_{\bp} d_{\bp}^{\dagger}}$.
   There will be $n_{\bp}$ such contractions, each of which contribute $(-2\lambda/N)$ to $\avg{ \of{-1}^{n_{\bx}} }$.  Summing over the momenta $\bp$, we see
\begin{eqnarray}\label{avg}
 F \of{1} = - \frac{2}{N} \sum_{\bp} \avg{n_{\bp}} = -2 \avg{n}.
\end{eqnarray}
  
At second order, there are two classes of nonzero terms; we can either choose two contractions of equal momenta, $f^{(1)}_{{\bp\bk}}=\overline{c_{\bp} d_{\bp}^{\dagger}} \times \overline{c_{\bk} d_{\bk}^{\dagger}}$ or we can contract two pairs of unequal momenta in the form $f^{(2)}_{{\bp\bk}}=\overline{c_{\bp} d_{\bk}^{\dagger}} \times \overline{c_{\bk} d_{\bp}^{\dagger}}$. In both these cases, all the momentum dependent phases from the exponentials in (\ref{avg}) cancel. This is a consequence of translational invariance, and will be true at each order.  
As long as $\bp\neq\bk$, each of these two terms contribute the same amount: $f^{(1)}_{{\bp\bk}}=f^{(2)}_{{\bp\bk}}=n_\bp n_\bk (2\lambda/N)^2$.  The case where $\bp=\bk$ can be ignored in the thermodynamic limit, as the sum of their contribution will scale as $1/N$.  
 We  sum over $\bk$ and $\bp$, dividing by 2 to take care of overcounting, to arrive at:
\begin{eqnarray}
F \of{2} = 4 \avg{n}^{2}.
\end{eqnarray}

\begin{figure}
\includegraphics[width=3.3in]{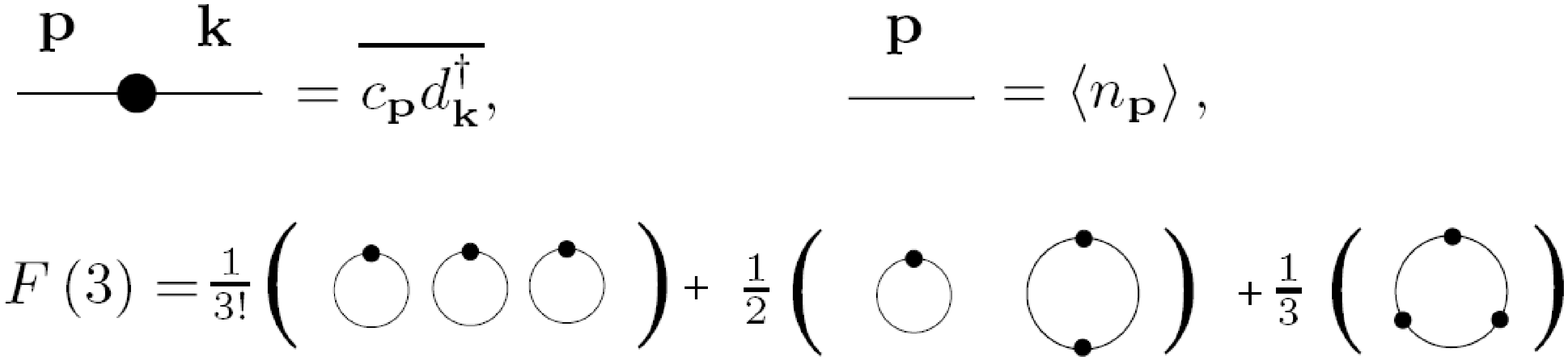}
\caption{Diagrammatic representation of the expansion in $\lambda$, with the third order term shown explicitly. The cluster decomposition principle applies once the terms are grouped into rings as described in the proof, and the combinatorics of the summations over momenta lead to the exponentiation of all nonvanishing diagrams. The numerical contribution of $F \of{3}$ comes from summing over all momenta for each leg in the sum of ring diagram, taking into account the combinatorics described in the text.}\label{diagrams}
\end{figure}

Generalizing to the case of of $m$ contractions is best performed by introducing a diagramatic language and reorganizing.  A contraction $\overline{c_{\bk} d_{\bp}^{\dagger}}$ is represented by a vertex.  Each vertex has two lines, representing the momenta $\bk$ and $\bp$.  Vertices which are connected share the same momentum.
$\lambda^m F(m)$ is given by a sum of all possible closed configurations of $m$ vertices, attributing a factor $-2\lambda/N$ to each vertex, and a factor $\langle n_\bp\rangle$ to each line.  Any given diagram will have associated combinatoric factors to eliminate double-counting.  
The result for $F(1)$ is just a single loop.  The two contributions to $F(2)$, are the product of single loops, and the ring diagram with two vertices.  The diagrams for $F(3)$ are also shown in Fig.~\ref{diagrams}, yielding $F(3)=-8 \langle n\rangle^3$.

In any term where there is a product of $m$ identical diagrams, there will be a combinatoric factor of $\of{1/m!}$.  This coefficient represents the fact that permuting the different diagrams does not produce new terms.  This property implies that the linked cluster expansion applies, and upon exponentiating, disconnected diagrams factor out and can be discarded. Therefore,
$\sum_m \lambda^m F(m)=\exp(\sum_m \lambda^m R(m))$, where $R(m)=C_m (-2\lambda/N)^m \langle n\rangle^m$, is the contribution from the linked diagram with $m$ vertices: the $m$-ring.  The combinatoric factor $C_m=1/m$ results from the $m$ different starting position of the ring.  It can also be thought of as the ratio of the $(m-1)!$ ways of arranging $m$ momenta into a ring, divided by the $m!$ different possible permutations of those momenta.

Since each $m$-ring amplitude scales as the $m$th power of the average density, our end result is a series in the density of the system. Summing the series yields
%
\begin{eqnarray}
\avg{ \of{-1}^{n_{\bx}} } = \prod_{m=1}^{\infty} e^{R_{m}} = e^{ \sum_{m=1}^{\infty}R_{m} } = \frac{1}{1+ 2 \avg{n}}.
\end{eqnarray}
In terms of our observable operators:
\begin{eqnarray}
\avg{P_{\bx}} = \frac{1 - \avg{ \of{-1}^{n_{\bx}} }}{2} = \frac{\avg{n}}{1+2 \avg{n}}.
\end{eqnarray}
When $\avg{n} \ll 1$, one has the expected behavior that $\avg{n_{\bx}} = \avg{P_{\bx}}$.  As $\avg{n}\rightarrow \infty$, there is no bias towards even or odd occupations and $\avg{P_{\bx}}\to 1/2$.
%

\subsubsection{Two-Site Expectation Value and Correlation Functions}
We now want to evaluate $\avg{ \of{-1}^{n_{\bx}} \of{-1}^{n_{\mathbf{0}}} }$. All of the combinatorical arguments of the previous section still hold, meaning that the expectation value can still be written as the exponential of the sum of ring amplitudes. However, beyond the 1-ring amplitude, the momentum dependent phases will no longer cancel, making the ring amplitudes depend on $\bx$. Adopting the conventions above, we make the following replacements:
\begin{eqnarray}
d_{\bk}^{\dagger} \rightarrow \of{ c_{\bk}^{\dagger} - \frac{2 \lambda}{\sqrt{N}} c_{\mathbf{0}}^{\dagger} - \frac{2 \lambda}{\sqrt{N}} c_{\mathbf{\bx}}^{\dagger} e^{-i \bk \cdot \bx} }, \nonumber \\ \sqof{ c_{\bp} , d_{\bk}^{\dagger} } = \delta_{\bp \bk} - \frac{2 \lambda}{N} \of{1 + e^{i \of{\bp - \bk} \cdot x} }.
\end{eqnarray}
As in the previous section, $\lambda = 1$ is a formal expansion parameter. The diagram structure is identical to the previous section, but each vertex now contributes $(2\lambda/N)(1+e^{i \of{\bp - \bk} \cdot x})=(4\lambda/N)e^{i\of{\bp - \bk} \cdot x/2}\cos\left((\bk-\bp)\cdot \bx/2\right) $.  The ring diagrams can be written:
\begin{eqnarray}
R_{1} \of{\bx} &=& -4 \avg{n},  \\
R_{2} \of{\bx} &=&  \frac{8}{N^{2}} \sum_{\bk,\bp} \avg{ n_{\bk} n_{\bp} } C_{\bp-\bk}C_{\bk-\bp}
 \nonumber \\\nonumber
R_{m } \of{\bx} &=& \frac{\of{-4}^m}{m N^{m}} \sum_{\bk_{1} \cdots \bk_{m}} \avg{ n_{\bk_{1}} \cdots n_{\bk_{m}} } 
\\&&\nonumber\qquad\qquad
\times 
C_{\bk_{m}- \bk_{1}}\cdots C_{\bk_{m-1}- \bk_{m}}
\\\nonumber
C_{\bp} &=&\cos(\bp\cdot\bx/2)
%
\end{eqnarray}
We will integrate out one of the momenta in the expression for $R_M$, generate a recursion relationship.  In particular, by using basic trigonometric identities we may write
\begin{eqnarray}
f_{\bk\bq}&=&\nonumber
\sum_{\bq} \frac{\avg{ n_{\bq} }}{N}  
C_{\bk - \bq } C_{\bq - \bp }
=
\frac{\avg{n}}{2}  
C_{\bk - \bp}
+ \frac{G_{\bx}}{2} 
C_{\bk + \bp}
\end{eqnarray}
where the Greens function $G_{\bx}$ is defined by
\begin{eqnarray}
G_{\bx} &\equiv&\sum_{\bk} \frac{\avg{n_{\bk}}}{N}   e^{i \bk \cdot \bx} = \avg{c_{\bx}^{\dagger} c_{0}}.
\end{eqnarray}
To produce a closed set of recursion relationships, we introduce a modified ring amplitude
\begin{eqnarray}\label{aux}
W_{m} \of{\bx} &=& \of{\frac{-2}{N}}^{m} \frac{2^{m}}{m} \sum_{\bk_{1} ... \bk_{m}} \avg{ n_{\bk_{1}} \times ... n_{\bk_{m}} } \\ &&\times \nonumber
C_{\bk_{m}- \bk_{1}} \cdots C_{\bk_{m-2} -\bk_{m-1}}\times C_{\bk_{m-1} + \bk_{m}},
\end{eqnarray}
where the final cosine depends on the sum of two neighboring momenta rather than the difference. If one repeats the same procedure for Eq.~(\ref{aux}), integrating out $\bk_{m-1}$, one will find a term where two consecutive cosines depend on the sum of momenta.  By taking $\bk_{m-2}\to-\bk_{m-2}$, and using time reversal symmetry, one can express such a product in terms of $R_{m-1}$.
 The resulting recursion relations are
\begin{eqnarray}
R_{m} = -2 \avg{n} \frac{m-1}{m} R_{m-1} - 2 G_{\bx} \frac{m-1}{m} W_{m-1}, \nonumber \\
W_{m} = -2 G_{\bx} \frac{m-1}{m} R_{m-1} - 2\avg{n}  \frac{m-1}{m} W_{m-1}.
\end{eqnarray}
It is straightforward to solve these equations for $R_{m}\pm W_{m}$, to find
\begin{eqnarray}
R_m\pm W_m&=&(2/m)\left[-2\left(\langle n \rangle\pm G_\bx\right)\right]^m
\end{eqnarray}
where we have used that
$R_1=-4\langle n \rangle$ and $W_1=-4 G_\bx$. This yields the correlation function
\begin{eqnarray}
\sum_{m = 1}^{\infty} R_{m} = - \log \sqof{\of{1  + 2 \avg{n} }^{2} - 4 G_{\bx}^{2} }, \\
\avg{ \of{-1}^{n_{\bx}} \of{-1}^{n_{0}} } = \frac{1}{ \of{1  + 2 \avg{n} }^{2} - 4 G_{\bx}^{2} }.
\end{eqnarray}
This allows us to express the correlator in Eq.~(\ref{defd}) as
\begin{eqnarray}\label{ans1}
D_{\bx \by} = \frac{ G_{\bx-\by}^{2} }{ \of{1+ 2 \avg{n}}^{2} \of{ \of{1+ 2 \avg{n}}^{2}-4 G_{\bx-\by}^{2}    }   }.
\end{eqnarray}
For the non-interacting gas, the density-density correlation function is simply the product of Greens functions,
\begin{eqnarray}
\avg{n_{\bx} n_{\mathbf{0}}} - \avg{n_{\bx}} \avg{n_{\mathbf{0}}} &=& \nonumber\frac{1}{N^{2}}\sum_{\bk,\bp} \avg{c_{\bk}^{\dagger} c_{\bp} c_{\bp}^{\dagger} c_{\bk}} e^{i \of{\bk - \bp} \cdot \bx}\\ &=& G_{\bx}^{2},
\end{eqnarray}
Thus the parity correlation function is only a function of the average density $\avg{n}$ and the density-density correlation function.  The denominator in Eq.~(\ref{ans1}) is always larger than 1, implying that the parity correlations are always smaller than the density-density correlations.  In particular, 
at large densities, 
$D_{\bx \by} \sim \avg{n}^{-2}$.  The characteristic length scale of these correlations, is however set by the density-density correlations.

\subsubsection{Bose Condensed Phase}\label{sf}
The previous formalism can be modified slightly to describe the Bose-Condensed phase, where there is macroscopic occupation of the $k=0$ mode.  We work in an ensemble where the number of condensed particles $n_0$ is fixed and extensive, so
\begin{figure}[tbp!]
\includegraphics[width=3in]{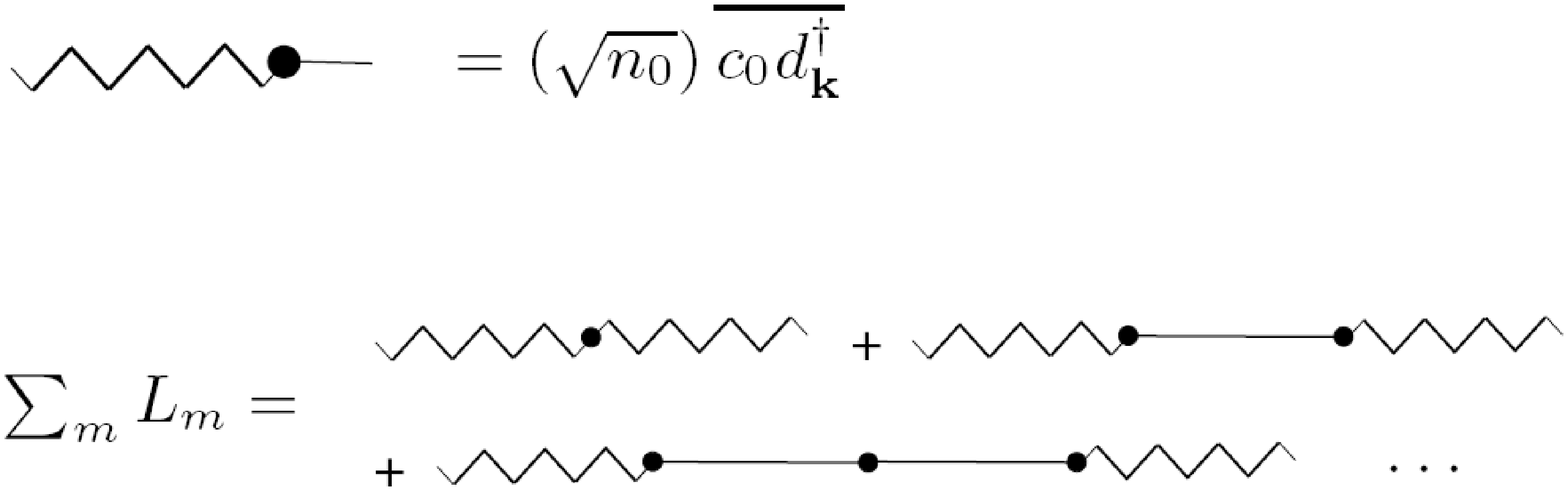}

\caption{Top: new vertex and line needed to include condensate.  Bottom: Contribution to $\langle(-1)^{n}\rangle=\exp(\sum_m R_m+L_m)$.}
%
\label{sfdiagrams}

\end{figure}
\begin{widetext}
\begin{eqnarray}
\avg{ \of{-1}^{n_{\bx}} } 
= Z^{-1} \sum_{\{n_k\}} \frac{1}{n_0! \prod_{i = 1}^{M} n_{i} !} \left< 0 \right| \of{c_0}^{n_0}
 \of{c_{\bk_{1}}}^{n_{1}} 
\cdots \of{c_{\bk_{M}}}^{n_{M}} 
 \of{-1}^{n_{\bx}} \of{c_{\bk_{1}}^{\dagger}}^{n_{1}}  
\cdots \of{c_{\bk_{M}}^{\dagger}}^{n_{M}} 
 \of{c_0^\dagger}^{n_0}
\left| 0 \right> e^{-\sum_k \frac{E_k n_k}{T}}
\end{eqnarray}
\end{widetext}
where all sums over $k$ do not include $k=0$.

One readily includes the extensivity of $n_0$ through a modification of our diagramatic language. To motivate our formalism, we first consider the zero temperature expectation value
\begin{eqnarray}\label{zeroT}
\avg{ \of{-1}^{n_{\bx}} } &=& \langle 0| \of{c_0}^{n_0} \of{d_{0}^{\dagger}}^{n_0}|0\rangle \nonumber \\ &=& \sum_{m=0}^{n_{0}} \frac{n_0 !}{\of{n_0 -m}! m!} \of{ - \frac{2 \lambda}{N} }^{m} \nonumber \\ &=& \of{1 - \frac{2\lambda}{N}}^{n_0}.
\end{eqnarray}
In the thermodynamic limit $(1-2\lambda/N)^{n_0}\to \exp(-2\lambda n_{\rm{s}})$, where the condensate density is $n_{\rm{s}}=n_0/N$.  This agrees with our argument in Sec.~\ref{ideal}.

In the thermodynamic limit ($n_0,N\to\infty$) this zero temperature gas has $F(m)=(-2\lambda/N)^m n_0!/(m! (n_0-m)!)\to (-2 \lambda n_{\rm{s}})^m/m!$.  The natural way to express this diagrammatically is to again have $m$ vertices, each of which contribute $(-2\lambda/N)$, and attach to each of them two short jagged lines, each of which contribute $\sqrt{n_0}$.  These jagged lines attach to only a single vertex, and in a coherent state formulation would have the meaning of the condensate order parameter $\langle a_0\rangle$  \cite{leggett}.  The denominator $m!$ represents the various permutations of the vertices.  

At non-zero temperature, one will have both contributions from contractions with non-zero $\bk$, described by the diagramatic language introduced in Sec.~\ref{finite}, and contractions with $\bk=0$.  Generically one will have ring diagrams (Fig.~\ref{diagrams}), and line diagrams, illustrated in Fig.~\ref{sfdiagrams}.  The diagrams describing Eq.~(\ref{zeroT}) are a special case of the line diagrams. 
We again exponentiate the series, finding 
$\langle(-1)^{n}\rangle=\exp(\sum_m R_m+L_m)$, where $R_m$ is the previously introduced $m$-vertex ring-diagram (with the $\bk=0$ mode removed) and $L_m$ is the $m$-vertex line diagram,
\begin{eqnarray}
L_{m} &=& n_0 \of{\frac{-2 \lambda}{N}}^m \sum_{\bk_1 ... \bk_m} n_{\bk_1} ... n_{\bk_m} \overline{c_0 d_{\bk_1}^{\dagger}}\,\overline{c_{\bk_1} d_{\bk_2}^{\dagger}} ... \overline{c_{\bk_m} d_{0}^{\dagger}} \nonumber \\
&=& n_{\rm{s}} \of{n_{\rm{ns}}}^{m} \of{-2 \lambda}^{m}, \nonumber
\end{eqnarray}
where $n_{\rm{ns}}=\langle n\rangle-n_{\rm{s}}$ is the density of non-condensed particles. 
Summing this series yields:
\begin{eqnarray}
\avg{ \of{-1}^{n_{\bx}} } 
&=& \frac{1}{1 + 2 n_{\rm{ns}} } \exp{ \frac{ -2 n_{\rm{s}} }{1 + 2 n_{\rm{ns}}} } .
\end{eqnarray}

A similar approach can be used to calculate $\avg{ \of{-1}^{n_{\bx}}\of{-1}^{n_{\mathbf{0}}} } $.  As before, one simply changes the vertex to account for the extra phase factors.  Our previous calculation of the ring diagrams goes through unchanged.  For the line diagrams we again produce a recursion relationship by integrating out one of the momenta.   
The result is that
\begin{eqnarray}
L_{m} &=& n_{\rm{s}} \of{ n_{\rm{ns}} + \bar{G}_\bx }^{m} \of{-2 \lambda}^{m},
\end{eqnarray}
where $\bar{G}_\bx=G_\bx-n_{\rm{s}}$.
 The correlator then becomes
\begin{eqnarray}
\avg{ \of{-1}^{n_{\bx}} \of{-1}^{n_{\mathbf{0}}} } &=& \frac{\exp \frac{  -4 n_{\rm{s}} }{(1 + 2 \of{  n_{\rm{ns}}  + \bar{G}_\bx } ) }}{ \of{1 + 2 n_{\rm{ns}}}^{2} - 4 \bar{G}_{\bx}^{2} }.
\end{eqnarray}
The even-odd correlation function will thus decay exponentially to zero with increasing superfluid density.

\subsubsection{Weak Interactions}\label{weak}

 The diagramatic series which we have developed may readily be expanded to allow for a perturbative treatment of interactions.  A new vertex, representing interparticle interactions will appear.  Physically, repulsive interactions will suppress number fluctuations, hence increase $\langle (-1)^{n_\bx}\rangle$.  One expects that when temperature is large compared to the interaction energy the correlation length will still be dominated by thermal effects, and our free particle results will be valid.  The lowest temperatures produced in experiments are a fraction of the bandwidth, and hence a fraction of the hopping $t$.  Thus when $t\gg U$ (where $U$ is the on-site interaction energy) our theory should suffice for describing the correlations.
  Moreover, a Feshbach resonance can always be used to reduce the importance of interactions.

\section{Large $U$ Bose-Hubbard Model}\label{largeU}
\subsection{Perturbation Theory about the Mott state}
We now consider the opposite limit, where interactions are strong compared to the hopping.  This is typically described by the Bose-Hubbard model \cite{bloch}:
\begin{eqnarray}
H_{bh} = \sum_{i} \of{U n_{i} \of{n_{i} - 1} - \mu_{i} n_{i} } - \sum_{ij} t_{ij} a_{i}^{\dagger} a_{j}
\end{eqnarray}
Here, $\mu_{i}$ is a site dependent potential that combines the bare chemical potential and any trapping potentials.  Here we will just treat the uniform system, taking $\mu_i=\mu$ -- as before we will use a local density approximation to treat the trap.  We want to calculate correlation functions of the form:
\begin{eqnarray}
D_{jk} \equiv \avg{ P_{j} P_{k} } - \avg{P_{j}} \avg{P_{k}}, \nonumber
\end{eqnarray}
where $P_{j}$ is the even-odd projector.

Following Freericks  \textit{et al}\cite{freericks}, and the related works of Eckardt \textit{et al}\cite{eckhardt,teichmann}, we perturbatively calculate the nearest-neighbor correlator in powers of $t/U$.  This expansion breaks down in the superfluid phase, but captures the leading order correlations in the Mott phases. The other typical approximation used to investigate the Bose-Hubbard model, the Gutzwiller ansatz, is not appropriate for calculating $D_{jk}$. By fiat, the Gutzwiller ansatz restricts correlations to be zero or infinite range. We will only work to second order in $t$, although the generalization to higher orders is straightforward.

Second order perturbation theory gives
\begin{align} \nonumber
\langle P_j\rangle&-\langle P_j\rangle_0 =
\int_0^\beta d\tau \int_0^\tau d\tau^\prime \langle
 H_{hop}(\tau)H_{hop}(\tau^\prime)P_j\rangle_{H_0}\\
 \label{pert}
 &-\int_0^\beta d\tau \int_0^\tau d\tau^\prime \langle H_{hop}(\tau)H_{hop}(\tau^\prime)\rangle_{H_0},
\end{align}
where $H_0=\sum_{i} \of{U n_{i} \of{n_{i} - 1} - \mu n_{i}} $ is the local part of $H$, and $H_{hop}=- \sum_{ij} t_{ij} a_{i}^{\dagger} a_{j}$.  The expectation value in the local ensemble is 
\begin{equation}
\langle \hat X\rangle_{H_0}=\frac{{\rm Tr} \: e^{-\beta H_0} \hat X}{{\rm Tr} \: e^{-\beta H_0}},
\end{equation}
and the interaction picture operators $H_{hop}(\tau)=e^{-H_0\tau} H_{hop} e^{H_0\tau}$.  An expression similar to (\ref{pert}) describes $\langle P_j P_k\rangle$.

\begin{figure}
\includegraphics[width=\columnwidth]{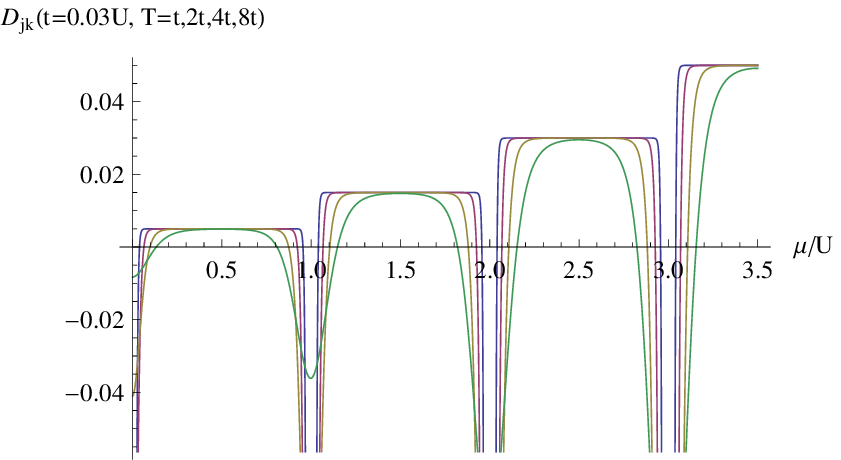}

\caption{(Color Online) Leading order strong coupling perturbation theory calculation of nearest neighbor parity correlations $D_{jk}$ at temperatures $T=\cuof{t,2t,4t,8t}$ (from top to bottom in the plateaux; blue, purple, yellow, green in color versions of this article) for $t=0.03U$ as a function of $\mu/U$. At small $t/U$ these are positive in the Mott state, and negative in the superfluid.  The perturbation theory breaks down in the superfluid, but the sign of the correlations should be robust.}\label{djk}
\end{figure}

In Eq.~(\ref{pert}), we introduce a resolution of the identity between each operator, consisting of the states with a fixed number of particles on every site.  As detailed in \cite{freericks} the resulting expression factors, leaving only chains of hopping operators which pass through site $j$ (or $k$ in the case of calculation $\langle P_j P_k\rangle$).  The resulting expressions are most conveniently written in terms of the local Greens functions
%
\begin{eqnarray}
G_{j} \of{\tau,\tau' } \equiv - \avg{ T_{\tau} a_{j} \of{\tau} a_{j}^{\dagger} \of{\tau'}          }_{H_{0}}, \\
G_{j}^{P} \of{\tau,\tau' } \equiv - \avg{ T_{\tau} a_{j} \of{\tau} a_{j}^{\dagger} \of{\tau'}          P_{j} }_{H_{0} },
\end{eqnarray}
where the time ordering operator $T_\tau$ places operators at earlier imaginary times towards the right.  Implicitly we set the imaginary time of the $P$ operators to zero so that they always appear on the right.
We define the bare single site energies, $\epsilon_{n}=\of{U n \of{n - 1} - \mu n}$, and the single site partition function $Z_{ss}=\sum_n e^{-\beta \epsilon_n}$.  The Greens functions are then explicitly given as 
\begin{eqnarray}
G_{j} \of{\tau , \tau' } = \sum_{n} \rho_{ n} \sqof{ \of{n+1} \theta \of{\tau - \tau'} e^{ \of{\tau' - \tau} \epsilon_{ n}^{+} } } \nonumber \\  + \sum_{n} \rho_{ n} \sqof{ n \theta \of{\tau' - \tau} e^{\of{ \tau- \tau'} \epsilon_{ n}^{-} }       }.
\end{eqnarray}
where
$\rho_{n} \equiv e^{-\beta \epsilon_{n}}/Z_{ss}$, and $\epsilon_n^{\pm}=\epsilon_{n\pm1}-\epsilon_n$.
$G_{j}^{P} \of{\tau, \tau'}$ is simply given by removing the even $n$ terms in the expression for $G_{j}$. One then finds
\begin{widetext}
\begin{eqnarray}
\avg{P_j}-\avg{P_j}_0&=& t^{2} \sum_{l \neq j} \of{ X_{jl}^{P} - X_{jl} }, \nonumber \\
\avg{P_j P_k} - \avg{P_j P_k}_0 &=& t^{2} \sqof{ X_{jk}^{PP} + \sum_{l \neq k,j} \of{ \avg{P_{j}}_{0} X_{kl}^{P} + \avg{P_{k}}_{0} X_{jl}^{P} } - \avg{P_{j}}_{0} \avg{P_{k}}_{0} \of{X_{jk} + \sum_{l \neq k,j} \of{X_{jl} + X_{kl}} }  } \nonumber
 \\
X_{jk} &=& \int_{0}^{\beta} \!\!dx \int_{0}^{\beta}\!\!dy\, G_{k} \of{x,y} G_{j} \of{y,x} = \frac{1}{2 U^{2}} \sum_{n,m}^{\infty} \rho_{n} \rho_{m} \of{\Gamma_{nm} + \Gamma_{mn}}, \nonumber \\
X_{jk}^{P} &=& \int_{0}^{\beta} \!\!dx \int_{0}^{\beta}\!\!dy\, G_{k}^{P} \of{x,y} G_{j} \of{y,x} = \sum_{n,m}^{\infty} \rho_{n} \rho_{m} P \of{n} \of{\Gamma_{nm} + \Gamma_{mn}}, \nonumber \\
X_{jk}^{PP} &=& \int_{0}^{\beta} \!\!dx \int_{0}^{\beta}\!\!dy\, G_{k}^{P} \of{x,y} G_{j}^{P} \of{y,x} = \sum_{n,m}^{\infty} \rho_{n} \rho_{m} P \of{n} P \of{m} \of{\Gamma_{nm} + \Gamma_{mn}}, \nonumber \\
\Gamma_{nm} &=& \frac{ \of{1+m}n \of{-1 + \beta U \of{1-m+n}+e^{\beta U \of{-1-m+n}} }}{\of{1+m-n}^{2}  }, \nonumber
\end{eqnarray}
\end{widetext}
As shown, the integrals reduce to a rapidly converging double sum which we compute numerically.  We plot these correlation functions in Fig.~\ref{djk}.
The dominant feature is a series of plateaus where $D_{jk}$ is positive, punctuated by downward spikes where $D_{jk}$ becomes negative.  The size of the plateaus increase monotonically as $\mu$ increases.

Both the plateaus and the spikes can be qualitatively understood from a truncated two-site model.  Deep in the $m$-particle Mott regime, one expects that configurations at two neighboring sites would be dominated by the states $\cuof{\left| m,m \right>, \left| m-1,m+1 \right>, \left| m+1,m-1 \right>}$, where the two integers represent the number of particles on each site.  The probability of being in either of the latter two states is $b\sim m (t/U)^2$, where the factor of $m$ comes from Bose enhancement.  The correlations function $\langle P_j P_k\rangle-\langle P_j\rangle \langle P_k\rangle = 4 (b-b^2)$ is positive, reflecting the fact that the parity of two neighboring sites is correlated.

Conversely, near the single-site level crossings ($\mu/U=m$), the configurations of neighboring sites are dominated by  $\cuof{\left| m,m \right>,\left| m,m+1 \right>,\left| m+1,m \right>}$.  At finite $t$ the latter two states are energetically favorable, resulting in anticorrelations between the parity at the two sites.  Although in this regime our perturbation theory breaks down, the anticorrelations should be robust.

\section{Spin-$\frac{1}{2}$ Fermions}\label{ferm}

We readily calculate the parity correlations of spin-1/2 fermions.  Since each site can have at most one particle of each spin, the parity is
\begin{equation}
P_i = n_{i\uparrow}+n_{i\downarrow}-2 n_{i\uparrow} n_{i\downarrow},
\end{equation}
and the parity correlations can be readily expressed in terms of density correlations.  In the noninteracting limit, these factor, yielding for $\bx\neq0$
%
\begin{eqnarray}
\avg{P_{\bx} P_{0}} - \avg{P_{\bx}} \avg{P_{0}} = 4 G_{\bx}^{4} - 2 \of{1 - \bar n/2}^{2} G_{\bx}^{2},
\end{eqnarray}
where $G_{\bx}= \frac{1}{N_{s}} \sum_{\bk} \avg{n_{\bk}} e^{i \bk \cdot \bx}$ is the single particle density matrix. The correlations vanish when $\bar n=0,2$.  For intermediate $\bar n$ the correlations can be positive or negative at short distances, but are always negative at large distances.

\section{Conclusion}

We have calculated the atomic parity correlation functions $D_{\bx \by}$ for three of the most common systems in optical lattice physics, non-interacting bosons, the Bose-Hubbard model, and non-interacting fermions. In the free boson system, we found that $D_{\bx \by}$ decays to zero at long distances at the same rate as the square of the thermal Green's function, matching the behaviour of the  density-density correlation function. In addition, the magnitude of $D_{\bx \by}$ decays to zero with increasing density, and with increasing condensate density.   In the Bose-Hubbard model, we demonstrated a simple link between the parity and density operators, and calculated the parity correlation functions to second order in $t/U$ using imaginary time perturbation theory.

In fermionic systems, the even-odd correlators are calculated in the same manner as the density-density correlation functions, and no new theoretical machinery is needed to interpret them.

\section{Acknowledgements}

Eliot Kapit would like to thank Kaden Hazzard for useful conversations related to this work.  
This work was supported by a grant from the Army Research Office with funding from the DARPA OLE program, and by the Department of Defense (DoD) through the National Defense Science \& Engineering Graduate Fellowship (NDSEG) Program.

\end{document}